\documentclass[sigplan,screen,authorversion,nonacm]{acmart} 

\usepackage{times}
\usepackage{url}
\usepackage{listings, lstautogobble}
\usepackage{xspace}

\usepackage[nameinlink]{cleveref}
\usepackage{hyperref}

\usepackage{balance}

\newcommand{\bosque}{\textsc{Bosque}\xspace}
\newcommand{\bapi}{\textsc{BAPI}\xspace}

\newcommand{\tecton}{\textsc{Tecton}\xspace}

\newcommand{\eg}{\hbox{\emph{e.g.}}\xspace}

\newcommand{\etc}{\hbox{\emph{etc.}}\xspace}

\newcommand{\cf}[1]{\texttt{#1}}

\definecolor{cgreen}{rgb}{0.25,0.5,0.35} 
\definecolor{stringred}{rgb}{0.6,0,0} 

\lstdefinelanguage{bosque}{
keywords={api, function, entity, field, requires, ensures, invariant, abort, env, permissions, example, type, \$events, \$return, \$src, agent, let, if, none, return},
keywordstyle=\color{blue}\bfseries,
identifierstyle=\color{black},
alsoother={@},
sensitive=true,
morecomment=[l]{\%\%},
morecomment=[s]{\%**}{**\%},
commentstyle=\bfseries\color{cgreen}\ttfamily,
}
 
\setcopyright{cc}
\setcctype{by}
\copyrightyear{2026}
\acmYear{2026}
\acmConference[SpecOps '26]{1st International Workshop on Specification-Driven Development Life Cycle}{October 04--09, 2026}{Oakland, CA, USA}
\acmBooktitle{1st International Workshop on Specification-Driven Development Life Cycle (SpecOps '26), October 04--09, 2026, Oakland, CA, USA}
\acmDOI{10.1145/3842652.3843194}
\acmISBN{979-8-4007-2970-6/2026/10}



\begin{document}

\title{A-JIT: Agentic Just-In-Time Software Construction}

\author{Mark Marron}
\affiliation{%
  \institution{University of Kentucky}
  \city{}
  \country{}
}
\email{mark.marron@uky.edu}

\author{Earl T. Barr}
\affiliation{%
   \institution{University College London}
   \city{}
   \country{}
}
\email{e.barr@ucl.ac.uk}

\begin{abstract}
Traditional software delivery assumes a static paradigm: code is constructed prior to execution and 
deployed as a fixed artifact. We present Agentic Just-In-Time Software Construction (\emph{A-JIT}), 
a paradigm that replaces static binaries with dynamic, software systems that can perpetually evolve 
to meet changing demands. In \emph{A-JIT}, an application is an integrated assembly comprising code, 
a runtime harness, and an embedded AI agent that continuously observes system usage and live execution 
traces. Much like a traditional JIT compiler specializes machine code to runtime execution paths, 
\emph{A-JIT} specializes software logic, workflows, and tool interfaces to meet the specific needs of 
the end-user. By integrating synthesis directly into the ambient application lifecycle, \emph{A-JIT} 
enables applications to dynamically construct missing implementations, generate new capabilities on 
the fly, and continuously adapt to end-user behavior. We demonstrate how this model supports trace-driven 
human-AI co-construction and opens a new design space for adaptive, self-evolving software.
\end{abstract}

\maketitle

\section{Introduction}
Just-in-Time (JIT) compilation represented a major shift in thinking about the role of the compiler. 
Instead of pre-compiling all code before execution the JIT compiler integrates the compiler into 
the runtime environment, and then, compiles code just-in-time as it is needed during execution. We 
believe that the same shift in thinking can be applied to the software development process itself. 
Instead of writing code before execution, we can integrate the software development process 
into the runtime environment, and then, construct code just-in-time as it is needed during execution.

This paper introduces the concept of agentic just-in-time assisted application construction (\emph{A-JIT}). 
The concept depends on a combination of new language, runtime support, and agentic generation capabilities
to enable a new form of interactive software development:
\begin{itemize}
    \item Programming language support for explicitly denotating which parts of the program are intended 
    to be constructed just-in-time. 
    \item A runtime environment that can support the just-in-time construction of code, including the ability 
    to run and test code as it is being constructed.
    \item A novel conception of what an implementation is, allowing a human (or an AI agent) to construct a 
    complete code artifact or a partial solution for a specific set of traces.
    \item Integration with a fully tactile runtime value representation, allowing a human (or AI agent) to 
    directly author, exchange, and hot-load any program value.
\end{itemize}

\begin{figure}[t]
\centering
\begin{lstlisting}[language=bosque,xleftmargin=0.04\linewidth]
type Fahrenheit = Int; type Inch = Float;

type Celsius = Int; type MilliMeter = Float;

entity TempRange<T> { 
    field low: T; 
    field high: T;
    
    invariant $low <= $high;
}

entity Forecast {
    field temperature: TempRange<Fahrenheit>;
    field precipitation: Inch;
}

entity ForecastSI { 
    field temp: TempRange<Celsius>;
    field precip: Option<MilliMeter> = none;
}

function convert(forecast: Forecast): ForecastSI {
    abort; %%Not Implemented
}
\end{lstlisting}
\caption{Running example of a weather forecasts and conversion to SI format.}
\label{fig:runningexample}
\end{figure}

\paragraph{Developer Centric A-JIT}
We begin by exploring how these key features come together in practice and allow use to build a developer-driven 
A-JIT application.

The example code in \Cref{fig:runningexample} shows sample code involving weather forecasts. 
In this example, the code defaults to Imperial units for the \cf{temperature} and \cf{precipitation} fields.
However, other parts of the code need to work with SI units and have a slightly different definition where the 
field name (\cf{precip}) differs and is optional.

This requires the implementation of a mostly simple but non-trival conversion function that translates a \cf{Forecast} 
value, handling the differences in units and determining how to handle the optional nature of the precipitation field 
--- \eg ``Does \cf{0<Inch>} translate to \cf{some(0<MilliMeter>)} or \cf{none}?''\footnote{This choice can have substantial 
implications for interop with external systems and/or if the data needs to stored or transmitted in a format like JSON 
where the presence or absence of fields can have semantic meaning and representation costs.} in the SI representation?
As in many cases today, a developer may decide to defer the work to implement this function to focus on other logic 
in the application and simply write an \cf{abort} statement with a TODO note.

\paragraph{Automatic User-Driven A-JIT} Human developers, however, are not strictly required; rather, they represent just one point along a spectrum of human involvement in the A-JIT process.

In the more general scenario we envision the system autonomously observing user interactions and operating as a 
core feature of the system. Thus, it can be used by agentic AI systems, as a tool or part of a harness, or autonomously 
triggering proactive adaptations based on observed user behavior.

For example, a user may prompt a spreadsheet assistant agent to ``select all the rows where the submission has not been 
received by August 24''. This type of natural language query is a common situation where part of the command is clear and unambiguous, 
but other parts may require interpretation or additional context to execute correctly -- specifically is the date 
constraint inclusive, exclusive, AOE, local time, \etc? In today's systems, an LLM based agent will simply make some 
assumption based on some internal probability distribution. However, in an A-JIT system, the agent has a mechnism to 
explicitly inject an operation that will trigger clarification or adaptation when needed. 

This example is just one illustration of the potential for how an A-JIT system can enhance user interactions and 
system adaptability by integrating ambient monitoring, responsive, and synthesis directly into the application 
lifecycle. This capability blurs the boundary between developer, any AI agent, and end-user. In fact a developer 
might just code the broad strokes of the application, leaving the finer details (such as user preferences for flight 
sort order) and adaptations to be dynamically handled by the A-JIT system.
Conversely, developers may simply oversee the process, providing an initial skeleton and letting the system automatically generate the rest of the application based on user interaction traces.

\section{Code Holes}
\label{sec:holes}
A key capability for an A-JIT system is the ability to explicitly indicate that a particular task or piece of code is incomplete and needs to be filled in later.
Instead of relying on an ad-hoc set of conventions of \cf{TODO} comments, asserts, and issue numbers, the \bosque programming language~\cite{bosque} provides a specialized syntax and semantics 
as first-class parts of the language for just this purpose. Its \cf{hole} construct allows developers and agents to explicitly indicate that a particular task 
or piece of code is incomplete and needs to be filled in later, \emph{and}, decorate this incomplete task with additional metadata, such as the expected input 
and output types, document comments, pre/post conditions, and even information on where relevant input/output examples can be located.

The code in \Cref{fig:holes} shows an example of using hole expressions in the \cf{convert} function. In \bosque, this is as simple as using the \cf{?\_} construct 
for the implementation body. Additional metadata can be provided using the \cf{ensures} clause to specify an additional postcondition --- that a zero value in the standard 
Forecast should be mapped to \cf{none} in the SI version. The \emph{doc comment} for the function provides a natural place for normative documentation on the intended behavior 
of the function, and can be used by an agent to guide the synthesis of the missing implementation.

\begin{figure}[t]
\centering
\begin{lstlisting}[language=bosque,basicstyle=\scriptsize\ttfamily]
%** Convert from from standard Imperial units to SI. **%
function convert(forecast: Forecast): ForecastSI
    ensures forecast.precipitation == 0.0f ==> 
            $return.precip == none;
{
  ?_;
}
\end{lstlisting}
\caption{Example of using a hole expression in the \cf{convert} function to indicate that the implementation is incomplete and needs to be filled in later. 
The hole expression is decorated with an \cf{ensures} clause that specifies an additional postcondition -- that a zero value in the standard Forecast 
should be mapped to \cf{none} in the SI version.}
\label{fig:holes}
\end{figure}

In addition to traditional pre/post conditions and normative descriptions of the intended behavior, the \cf{hole} construct also supports linking to 
test cases or example input/output pairs that can be used to guide the synthesis of the missing implementation. These IO pairs play two critical roles 
in the A-JIT process. First, they provide a concrete specification of the expected behavior of the missing implementation, which can be used by an agent to
synthesize the missing code. More importantly, they provide a key point to provide runtime hooking, enabling dynamic behavior (\Cref{sec:enduserexperience}), and 
are the entry points where agents can step in mid-flight to synthesize custom logic (or outputs) independently or in collaboration with an end-user 
(\Cref{sec:partialandcomplete}).

Treating deferred implementation as a first-class language feature in A-JIT is critical to formalizing the workflow, enabling tool integration, and training architecture-focused AI agents.
This makes it both amenable to other automated tooling and provides a path for developing a corpus of ``tasteful'' examples for training future 
architecture focused AI agents.

\section{Partial and Complete Implementation}
\label{sec:partialandcomplete}

Afshari \emph{et al.} introduced the concept of lazy program completion~\cite{prorogued}; it is a key feature of the A-JIT system. The idea is that instead of requiring a complete 
implementation of a function or method, we can allow a partial implementation to be provided, and then, the system can automatically complete the implementation as needed. 
It allows a developer to run and experiment with an application even when some parts of the code are incomplete. 

A major limitation of this prior work is that the completion mechanism required a developer to hand-craft result values whenever a partial implementation was encountered. 
The effort required to manually construct these values is a major barrier to adoption and limits the utility of the approach. However, the advent of LLMs and, recently, 
the ability to construct and validate values that are consistent with the expected semantics of external APIs changes this calculus. The Tecton value generation 
framework~\cite{tecton}, initially developed to automatically generate test and mock values, can be repurposed in the A-JIT system to play the role of the human developer 
and automatically construct concrete output values for any inputs to a partial implementation.

In our running example (\Cref{fig:runningexample}), the A-JIT system will compile the application with the \cf{convert} function as special hook into the runtime analysis and 
generation system. When the code is run the A-JIT system will automatically detect that the \cf{convert} function is a partial implementation. First, the runtime will look 
for any available test cases or example input/output pairs that match the input and, if there is a match, will simply re-use the output value. Otherwise, it will trigger the 
agentic synthesis system to either 1) generate one, or more, concrete output values for the given input value directly using \tecton, or 2) use the existing set of 
constraints and input/output pairs to synthesize the missing implementation of the \cf{convert} function. 

In some cases, the system may not be able to generate a high-confidence concrete result or may find multiple plausible candidates. In these cases, the system can either 
prompt the user for additional guidance or, since \bosque supports fully branchable time-travel debugging, it could explore multiple plausible execution paths to attempt to find a valid result. 

\begin{figure}[t]
\centering
\begin{lstlisting}[language=bosque,basicstyle=\scriptsize\ttfamily]

%% Encountered input value for convert function
Forecast{
    TempRange<Fahrenheit>{ 58i, 80i },
    0.0f<Inch>
}

%% Tecton generated ForecastSI result value
ForecastSI{
    TempRange<Celsius>{ 
        14i, 
        27i
    }
}

%% A-JIT synthesized code for convert
function convert(forecast: Forecast): ForecastSI
    ensures forecast.precipitation == 0.0f ==> 
            $return.precip == none;
{
    var precip = none;
    if(forecast.precipitation != 0.0f) {
        precip = fromInchToMM(forecast.precipitation);
    }

    return ForecastSI{
        TempRange<Celsius>{ 
            fromFtoC(forecast.temperature.low), 
            fromFtoC(forecast.temperature.high)
        },
        precip
    };
}

\end{lstlisting}
\caption{Example of a Tecton generated ForecastSI result value for the convert function when given an input value. The system automatically detects that the convert function is a partial 
implementation and generates a concrete output value for the given input value using Tecton.}
\label{fig:convert-direct}
\end{figure}

In the example shown in \Cref{fig:convert-direct}, the A-JIT system automatically detects that the \cf{convert} function is a partial implementation and generates the correct concrete output 
value for the given input value using \tecton. A key feature of \tecton is the combination of the strong type and semantic constraints it leverages in \bosque sources and to 
run a combination of constrained decoding and assertion validation to ensure that the generated output value is consistent with the expected types and structures. 

Once the system has generated a sufficient number of IO pairs, it uses these pairs to synthesize a complete implementation of the \cf{convert} function (as shown in \Cref{fig:convert-direct}). 
The combination of a strong type system, semantic constraints, and a rich set of IO pairs allows an LLM agent to reliably construct a complete implementation of the \cf{convert} function that is 
consistent with the expected behavior. This model for the actual code generation is, in essence, a semantically well-founded formulation of the sub-agent architecture used in many agentic coding systems. 
This approach also ensures that, for any agent authored code, the ground truth of prompt data, examples, and constraints is always available as a source artifact that can be traced, versioned, 
and used to regenerate new implementations as needed.

\section{Tactile Values}
\label{sec:tactilevalues}
A key breakthough in enabling an A-JIT system to effectively adapt to user needs is the concept of tactile values. 
The baseline for these values is the ability to fully roundtrip any value into/out of a simple (and standard) 
human readable format. The \bapi format~\cite{bsqon} provides such a format and, in addition, has a rich set of 
standard types, like UUIDs and latitude/longitude, as well as allowing simple 
expression evaluations to be embedded directly within the tactile values themselves. This second aspect of 
particular utility when using these values to author the output of an (unimplemented) function when given an input value.

\begin{figure}[t]
\centering
\begin{lstlisting}[language=bosque,basicstyle=\scriptsize\ttfamily]

%% Sample Forecast value in BAPI format
Forecast{
    TempRange<Fahrenheit>{ 58i, 80i },
    0.0f<Inch>
}

%% Sample BAPI ForecastSI result value for convert
%% using expression evaluation 
ForecastSI{
    TempRange<Celsius>{ 
        fromFtoC($src.temperature.low), 
        fromFtoC($src.temperature.high)
    },
    none
}

%% Erroneous conversion result (accidental swap low/high)
%% Violates the validity conditions on TempRange
ForecastSI{
    TempRange<Celsius>{ 
        fromFtoC($src.temperature.high),
        fromFtoC($src.temperature.low)
    }
}

\end{lstlisting}
\caption{Examples of tactile values in BAPI format, including a sample Forecast value and a sample ForecastSI result value for the convert function using 
expression evaluation. An erroneous conversion result is also shown, which violates the validity conditions on TempRange.}
\label{fig:convert-literals}
\end{figure}

\Cref{fig:convert-literals} shows a sample Forecast input value in BAPI format and two sample ForecastSI result values for the convert function using expression evaluation. 
Three key features of these textual representations are that they are fully roundtrippable, they can be easily edited by a developer, and the underlying 
type information provides strong (checkable) type and structural constraints. 

The first two features allow a developer to easily author, edit, and test example values in the \bapi format. This allows a developer to quickly create a set of test cases and 
example input/output pairs that can be used to guide the synthesis of a missing implementation. The concise encoding that is possible in the \bapi format, as 
seen in the \cf{Forecast} example, makes these values easy to read and also reduces pressure on LLM context windows when working with these values to guide synthesis (\Cref{sec:partialandcomplete}). 
This ability is further enhanced by the ability to mix expressions and the presence of the implicit \cf{\$src} variable that allows the input value to be referenced directly 
in the output value. In the example this allows us to specify the output value in terms of the input value and the application on an existing \cf{fromFtoC} function. 

The third feature, the strong type and structural constraints, allows the system to validate the values and ensure that they are consistent with the expected types and structures.
This is critical in catching accidental inconsistencies in values, such as the erroneous conversion result shown in the example, where we have accidentally swapped the \cf{low} 
and \cf{high} fields in the ForecastSI result value. 

\section{The End-User Experience: Living, Self-Specializing Software}
\label{sec:enduserexperience}
The previous sections focused on a developer-centric view of the A-JIT system, where a developer is actively involved in the construction and adaptation of the application.
However, in the limit, the A-JIT system can be used to construct and adapt an application with minimal (or even without any) direct involvement from a human developer. In this section, 
we explore how the A-JIT system can be used to construct and adapt an application in response to end-user behavior and interactions. We consider the possibility using the A-JIT as a 
workflow monitor and then, as it identifies common workflows --- either per user or across a set of users --- it can proactively generate new \emph{robotic process application} (RPA) 
tools that are optimized for these workflows.

Consider a help desk ticketing system, where an operator may have some builtin automation to help them process tickets but, in many cases, 
must copy data, access various systems, and perform a series of steps that are not fully automated. In the A-JIT system model all of these steps can be monitored and traced by the 
embedded agent, and when a common workflow is identified, the agent can proactively merge a set of common traces to create a new streamlined workflow. Although conceptually simple, 
the task of identifying and clustering common workflows is a non-trivial problem, as often, related workflows diverge in parts of the handling. Thus, a direct trace-based synthesis of a 
new workflow often results in either a proliferation of overly specific workflows or a single overly general workflow. 

Instead the A-JIT system can experimentally use \bosque holes to inject hypothetical functions to handle divergence, argument calculation, and special case handling to abstract 
these types of differences away. Once it has one (or more) of these hypothetical functions in place, it can then use the same synthesis techniques to generate a complete implementation  
of these functions based either on the existing traces or running in shadow mode on new traces to validate the correctness of the generated implementation. The isolation and purity 
feature of \bosque~\cite{bosque} ensure that these generated functions are always safe to use and will not have any unintended side effects on the rest of the application.

This approach to customizing or extending an application by end-users introduces a number of novel UI/UX challenges. In particular, the system must provide a way for end-users to inspect 
and author data-results for missing implementations, and then, approve or reject the generated implementation. While developers may be comfortable with editing raw \bapi values, end-users may 
not be. Thus, the system must provide a set of UI/UX primitives, such as reactive UI models~\cite{peliui} or moldable patterns~\cite{NierstraszGirba2024Moldable}, that allow end-users to author 
and inspect these values in a more user-friendly way.

\section{Related Work}
Specification and requirements gathering is a major challenge for effective cooperation between human and AI agents in software development. 
Of particular interest in this space is prior work on multi-modal specification and interaction with \emph{End-User Programming}~\cite{agentpreicent,flashfill} systems, 
such as FlashFill~\cite{flashfill} and NLyze~\cite{nlyze}, and Prorogued Programming~\cite{prorogued}. The concept of holes in programming languages has a long 
history in synthesis~\cite{synthholes,nlyze} and type theory, with recent work on typed holes~\cite{Omar2019HazelnutLive,Omar2021Livelits} and moldable development 
patterns~\cite{NierstraszGirba2024Moldable}. A key challenge in this space is the ability to effectively specify and communicate the intended behavior of a program 
and to provide mechanisms to disambiguate and clarify the intended behavior~\cite{uist}.

Agentic monitoring and isolation have become critial topics with the increasing capability of LLMs and other AI agents to assist in software development. As agents 
use more tools and become more autonomous, the need for effective monitoring and isolation of agentic behavior becomes more important. Prior work on agentic monitoring and isolation
has focused on the use of sandboxing and isolation techniques to limit the scope of agentic behavior~\cite{agentsec1,agentsec2}. In the A-JIT model, the use of holes 
and a software isolation model allows for a more flexible and dynamic approach to monitoring and isolation.

\section{Onward!}
This paper outlines the vision for an agentic Just-In-Time Runtime, a 
software ecosystem designed to construct code just-in-time as it is needed during execution. 
We believe this approach to software development, and A-JIT, will play a critical role in addressing existing challenges 
in the software engineering lifecycle as well as emerging security and coding challenges presented by agentic systems. 

\section*{Data Availability}
All of the systems described in this paper are actively developed and publicly available via:\\ 
\url{https://github.com/BosqueLanguage/BosqueCore}.


\balance
\bibliographystyle{ACM-Reference-Format}
\bibliography{bibfile}


\end{document}